\documentclass[12pt, hidelinks]{article}
\usepackage[small, bf]{caption}
\usepackage[utf8]{inputenc}
\usepackage{fullpage}
\usepackage{appendix}
\usepackage{graphicx}
\usepackage{amsmath}
\usepackage{amssymb}
\usepackage{amsthm, amsfonts}
\usepackage{url}
\usepackage{mathtools}
\usepackage{chngcntr}
\usepackage{nicefrac}
\usepackage{hyperref}
\usepackage{subdepth}
\usepackage{forest}
\usepackage{tikzducks}
\usetikzlibrary{shapes.geometric,positioning}
\usetikzlibrary{decorations.pathmorphing,shapes}
\tikzset{elli/.style={ellipse,draw}}
\newsavebox\Car
\newsavebox\Tree

\usepackage{comment}
\usepackage{bbm}

\usepackage[style=alphabetic,backend=bibtex]{biblatex}
\bibliography{refs}

\theoremstyle{definition}

\theoremstyle{theorem}

\counterwithin*{lemma}{subsection}

\usepackage{tikz}
\usepackage{subcaption}
\usetikzlibrary{arrows,automata,calc}

\tikzset{
  treenode/.style = {align=center, inner sep=0pt, text centered,
    font=\sffamily},
  arn_n/.style = {treenode, circle, white, font=\sffamily\bfseries, draw=black,
    fill=black, text width=1.5em},
  arn_r/.style = {treenode, circle, red, draw=red, 
    text width=1.5em, very thick},
  arn_x/.style = {treenode, rectangle, draw=black,
    minimum width=0.5em, minimum height=0.5em}
}

\newcommand{\reals}{{\mbox{\bf R}}}

\newcommand{\naturals}{{\mbox{\bf N}}}

\newcommand{\Expect}{\mathop{\bf E{}}}

\newcommand{\argmax}{\mathop{\rm argmax}}

\newcommand{\eg}{{\it e.g.}}
\newcommand{\ie}{{\it i.e.}}

\newcommand{\BEAS}{\begin{eqnarray*}}
\newcommand{\EEAS}{\end{eqnarray*}}
\newcommand{\BEA}{\begin{eqnarray}}
\newcommand{\EEA}{\end{eqnarray}}
\newcommand{\BEQ}{\begin{equation}}
\newcommand{\EEQ}{\end{equation}}
\newcommand{\BIT}{\begin{itemize}}
\newcommand{\EIT}{\end{itemize}}

\newcommand{\kstar}{k^{\star}}

\title{An Analysis of Intent-Based Markets}
\author{
    Tarun Chitra\\ Gauntlet \\\texttt{\small tarun@gauntlet.xyz}
    \and Kshitij Kulkarni\\ UC Berkeley\\\texttt{\small ksk@eecs.berkeley.edu}
    \and Mallesh Pai\\ Rice University and SMG\\\texttt{\small mallesh.pai@mechanism.org}
    \and Theo Diamandis\\MIT\\ \texttt{\small tdiamand@mit.edu}
}

\begin{document}

\maketitle

\begin{abstract}
    Mechanisms for decentralized finance on blockchains suffer from various
    problems, including suboptimal price execution for users, latency, and a
    worse user experience compared to their centralized counterparts. Recently,
    off-chain marketplaces, colloquially called `intent markets,' have been
    proposed as a solution to these problems. In these markets, agents called
    \emph{solvers} compete to satisfy user orders, which may include complicated
    user-specified conditions. We provide two formal models of solvers'
    strategic behavior: one probabilistic and another deterministic. In our first model, solvers initially pay upfront costs to
    enter a Dutch auction to fill the user's order and then exert congestive, costly effort to search for
    prices for the user. Our results show that the costs incurred by solvers
    result in restricted entry in the market. Further, in the presence of costly
    effort and congestion, our results counter-intuitively show that a planner
    who aims to maximize user welfare may actually prefer to restrict entry,
    resulting in limited oligopoly. We then introduce an alternative,
    optimization-based deterministic model which corroborates these results. We
    conclude with extensions of our model to other auctions within blockchains
    and non-cryptocurrency applications, such as the US SEC's Proposal 615.
\end{abstract}

\section{Introduction}

Some of the most popular applications used on blockchains are decentralized exchanges, which allow users to trustlessly swap assets without a third-party intermediary.
Such exchanges have facilitated over \$2 trillion of transactions since 2018, and their usage continues to increase~\cite{defillama-dex}. 
However, decentralized exchanges (DEXs) still have worse user experience and price execution than centralized exchanges (CEXs), as they tend to be less capital efficient.
That is, it takes more passive capital to facilitate trades on a DEX than on a CEX.

Decreased capital efficiency on DEXs results from their reliance on continuous arbitrage to synchronize prices.
The most popular type of DEX, the Constant Function Market Maker (CFMM), was first shown to synchronize prices via arbitrage with an external market in 2019~\cite{angeris2019analysis, angeris2020improved}.
Recent work has demonstrated that this continuous arbitrage causes liquidity providers to face large losses and adverse selection costs, as their pricing and liquidity are completely public at all times~\cite{milionis2022quantifying, angeris2021replicating, cartea2023predictable}.
Moreover, the permissionless nature of CFMMs has empirically led to excess liquidity fragmentation~\cite{lehar2023liquidity} and maximal extractable value, or MEV, which does not have a precise analogue in the centralized world~\cite{xavier2023credible, kulkarni2022towards, daian2020flash}.
Despite these shortcomings, CFMMs still command substantial trading volume~\cite{the-block-dex}, and new mechanisms for improved capital efficiency, such as concentrated liquidity CFMMs~\cite{adams2021uniswap}, have narrowed the gap between decentralized and centralized exchanges.

\paragraph{Intent-based markets.}
Recently, there have been numerous proposals for settlement of on-chain transactions collectively known as \emph{intents} that claim to close the gap between centralized and decentralized exchanges.
Intent-based mechanisms allow users to specify precise conditions or covenants under which their transactions can be executed.
Transactions are then executed on a blockchain by third parties called \emph{solvers}.
Cryptographic primitives ensure that solvers cannot violate the user-specified conditions, and the validators of the blockchain verify that these conditions were satisfied.
CFMMs, conversely, deterministically quote a price based on public liquidity, and therefore, in contrast to intent-based markets, their execution price is known publicly.

Intents were first described by Anoma~\cite{yin_2023} as a way of credibly executing asynchronous transactions across multiple blockchains.These mechanisms generalize `request-for-quote' (RFQ) systems~\cite{0xrfq, o2021electronic, o2010quote}. In particular, intents allow users to express arbitrary covenants, whereas RFQ systems do not.
These covenants are guaranteed to hold throughout the execution of the transaction on multiple blockchains; a transaction is only settled after all blockchains have agreed on execution via a communication protocol.

Solvers are assumed to be rational economic actors who seek to satisfy a user's order only if it benefits them. Importantly, they hold or can acquire private information about the state of external markets that unsophisticated users cannot acquire. The central question is, therefore, whether a solver is incentivized to execute the user's order at `optimal' prices. In this sense, intent markets resemble principal-agent problems in which the user is a principal who employs an agent---the solver---to execute their order under the intent covenants.

\paragraph{Example intents.} To provide concrete examples of user intents, we consider a user who holds $1$ unit of token $T_1$ on blockchain $B_1$ (for example, the ETH token on the Ethereum blockchain). The user specifies an intent to convert the token into the maximum amount of output token $T_2$ on the same blockchain (for example, USDC on the Ethereum blockchain). We call this the \emph{maximum output} intent. As another example, consider the same user who now wants to move their token on blockchain $B_1$ to another blockchain $B_2$ (for example, from the Ethereum blockchain to the Solana blockchain) only if the realized variance of the token on a set of predefined venues on chain $B_2$ is less than a threshold. We call this the \emph{variance limited transfer} intent. 

While these two intents are superficially similar, only one can be easily verified after execution. In the first case, the validator is not able to verify that the user did, in fact, receive the maximum available output of token $T_2$ from all possible external venues. In the second case, however, a validator on chain $B_2$ is able to verify that the realized variance of the token did meet the conditions specified by the user. In this work, we do not focus on the issue of verifiability, but rather consider the economic incentives of solvers in fulfilling these intents.

\paragraph{Intents in practice: UniswapX.} Several intent-based marketplaces are already live in production in 2024. The first example we consider is UniswapX~\cite{Adams_2023}. 
Since its launch in late 2023, UniswapX has processed over \$3 billion in notional volume~\cite{UniswapX-dune}. This protocol incentivizes off-chain agents to provide users with better execution prices than those provided by CFMMs and, therefore, implements the maximum output transfer intent introduced earlier.
In UniswapX, solvers participate in a Dutch auction to provide price improvement over the price quoted by a CFMM. Solvers in UniswapX are referred to as `fillers'.
The solver who wins the auction then commits to filling the order at an execution price equal to their bid. The winning solver has to deliver the specified number of tokens within a fixed-time interval.
If the solver delivers the assets within the time-interval, the transaction is settled: the user receives the delivered assets less a payment to the solver for finding a better price.
Should the solver be delinquent and not deliver the assets, they are penalized and the user's trade is sent to a CFMM such as Uniswap. We will use this Dutch auction setup as a starting point for our analysis of solvers' incentives shortly.

\paragraph{Intents in practice: other protocols.} We now review several other intent protocols.  CoW Swap~\cite{Protocol_2023, cow-batch-auction, ramseyer2023speedex}, another intent-based protocol that implements the maximum output intent, has processed over \$30 billion since its inception in 2021~\cite{CoW_2023}. In CoW Swap, solvers participate in an auction to execute user orders at a uniform clearing price, also known as a `batch auction'. 

Another variant of an intent market is the so-called `oracle extractable value' market~\cite{Sengupta_2023, Lambur_2024, Benligiray_2023}.
These markets run auctions that sell the right to execute a trade immediately after a price update is sent on-chain via a price oracle~\cite{angeris2020improved, liu2021first}.
Such markets allow for decentralized finance protocols (who are the principals, akin to users in UniswapX) to outsource trade execution to agents and realize a welfare gain.

While there has been an immense amount of activity in building more general intent-based systems for purposes other than trading (\eg~SUAVE, Essential, and Anoma~\cite{Charbonneau_2023}), these systems do not currently have any formal guarantees. Further, the economic incentives of the participants in these systems have not been studied. 

\paragraph{Intents without blockchains.}
Another fulfillment model that is similar to intents but has been proposed for traditional equities markets is the US Security and Exchange Committee's Proposal 615~\cite{sec-prop-605}.
This rule aims to make Payment for Order Flow (PFOF) relationships more competitive.
These relationships, between brokerage firms, who purchase shares on behalf of their users, and wholesalers who execute transactions and settle trades for brokerages, have been controversial~\cite{angel2021gamestonk, bryzgalova2023retail, welch2022wisdom}. 

Currently, private agreements exist between brokerages and wholesalers that diminish competition, as noted in \cite{sec-prop-605}: ``Broker-dealers route more than 90\% of marketable orders of individual
investors in NMS stocks to a small group of six off-exchange dealers, often referred to as `wholesalers.'. The wholesaling business is highly concentrated...''
The SEC's rule proposes the construction of a Dutch auction wherein a wholesaler who receives orders from a brokerage must auction right to execute the orders at improved prices before fulfilling it themselves.
This extra auction is similar to blockchain intents like UniswapX or CoW Swap, but differs in the precise execution guarantees given to the user.
We discuss this auction and the applicability of our model to analyzing it in~\S\ref{sec:model-extensions}.

\paragraph{Competition.}
Despite the interest in intent-based systems, these systems are not necessarily guaranteed to perform significantly better than existing atomic mechanisms such as CFMMs.
Understanding conditions under which these systems provide improvements to existing mechanisms is therefore crucial to designing welfare-maximizing intent systems. Moreover, reliance on a network of solvers for fulfillment and execution in an intent market raises some questions: first, will the market remain competitive or concentrate to a few specialized actors? And second, how would such concentration affect user welfare?

\begin{figure}
    \centering
    \includegraphics[scale=0.32]{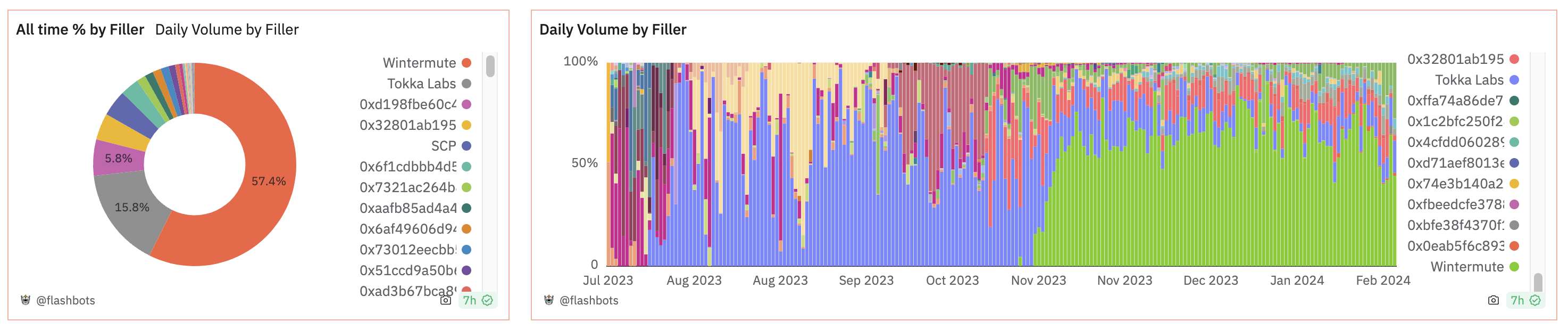}
    \caption{Evolution of competition within UniswapX over time~\cite{UniswapX-dune}. Left: the market share of different participants, in terms of the volume of orders filled. Right: the same data evolving over time. The data set has over 2,000 unique addresses that have participated, despite fewer than 20 addresses participating in January and Feburary 2024.}
    \label{fig:uniswap-x-figure}
\end{figure}

\paragraph{Empirical signatures of reduced competition in UniswapX.} 
In Figure~\ref{fig:uniswap-x-figure}, we see the evolution of competition within UniswapX \cite{UniswapX-dune}.
The left panel shows the percentage of volume processed by a particular solver and shows extreme concentration, with the top 3 solvers accounting for $\approx$79\% of daily volume transacted.
The right panel shows that the number of entrants into the auction has also decreased dramatically.
While over 2,000 unique addresses have participated as UniswapX fillers, with most of them participating in the earlier days of the protocol during Summer 2023, only 12 unique addresses participated in January 2024 (which can be viewed on the dashboard~\cite{UniswapX-dune}).
This concentration of fillers suggests that the UniswapX intent market suffers from a lack of competition, which may translate to user welfare losses.
We note that the filler role is currently permissioned by Uniswap Labs~\cite{Adams_2023}, so there is not free entry within this market.
However, the heavy concentration to 12 out of 2,000 participating addresses suggests that the market has barriers to competition even among the existing set of entrants.

\paragraph{This paper.}

In this paper, we formulate two simple models of solvers bidding in an auction
for a single user's intent to receive the best possible price when converting
one token into another. We aim to understand the solvers' economic incentives to
fulfill the user's intent. 

Our first model is probabilistic, relying on classical auction theoretic tools, where the standard auction setting is complicated by two features.
First, the
solvers initially face up-front \textit{entry costs}, which correspond to
infrastructure set-up costs that are endemic to any entity participating in
intent markets. Second, they face \textit{congestion costs} that are incurred
upon entry into the market---these costs potentially depend on the number of
other entrants, and are faced while competing with other solvers. For example,
the setting may require participants to exert costly effort to find liquidity to
satisfy the user's order in external markets. Further, participants may have to
invest in finding this liquidity prior to winning this auction, or indeed may
need to expend this effort just to know how much to bid.

We study the effects of these costs in the context of a Dutch auction. We start
by assuming there is a universe of $n$ possible solvers who may enter into the
auction. For each solver, entry costs are drawn i.i.d. from a distribution
$F_C$. As a function of these costs, an equilibrium number $k^*$ solvers enter
the auction, and then search for prices to satisfy the user's intent, which are
drawn i.i.d. from a distribution $F$. This search is affected by the congestion
costs and therefore, so is the solvers' bid in the Dutch auction. In this model, we have the following results:

First, in the presence of only entry costs but no congestion costs, we show
that the equilibrium number of entrants $k^*$ into the solver market is $O(\sqrt{n})$ when the solvers' prices are distributed exponentially, and $O(n^{1/3})$ when they are distributed uniformly on $[0,1]$. This means that the
percentage of potential solvers who participate in the auction goes to zero as
$n \rightarrow \infty$ since $\frac{\kstar}{n} \rightarrow 0$. We further show that for sufficiently fat-tailed price distributions, such as the Pareto distribution, one can
recover $\kstar = n$, although the user may not be able to realize large welfare increases due to significant bid shading.

Second, when congestion costs are present, the distribution of
prices is modified to a distribution $F(p, e)$, where $e$ is costly, congestive
effort exerted by the solver. Under congestive costs (such as those that might occur while searching for inventory in illiquid external markets), the effort that a solver exerts in
the search for better prices for the user degrades, and,
therefore, so does the user's welfare. In practice, this result suggests that
intent markets may provide welfare gains over CFMMs in highly liquid markets
where congestion costs are small, but they likely do not do so in markets for
extremely illiquid assets. Our results therefore connect costs faced by solvers
to the welfare gains that intent-based markets can promise to users. 

 Our second model takes a deterministic approach,
which is based on an optimization framework. This model naturally suggests the
Dutch auction as a means to solve the social welfare problem. Although it
introduces different modeling assumptions compared to our first model, the
deterministic model offers two advantages: it can be easily parameterized,
suggesting a fruitful avenue for future empirical work; and it likely has an
easier generalization to the multi-asset case, which we also leave to future
work.

Our analysis also implies several interesting features of intent markets. Entry costs may cause restricted entry into the markets. Further, when solvers are subject to congestion, limited oligopoly can
be better for user welfare than free entry; fewer entrants will have stronger
incentives to invest in higher effort due to less congestion. The welfare gains
for the user from additional investment from fewer entrants may outweigh the
losses to the user from increased bid-shading in an oligopolistic market.
Understanding the underlying costs faced by solvers is therefore important for an intent
platform to optimally design the market microstructure. 

\section{An auction-theoretic model}\label{sec:intent-single}

We begin with the simplest setting for an intent market: a single decentralized trade between two tokens. Consider a user who desires to swap $1$ unit of token $T_1$ into as many units of token $T_2$ as possible (this is an example of the maximum output intent introduced earlier). The user is faced with a choice to trade either with an infinitely liquid public market (an idealization of a CFMM) that quotes a price $p^*$, or with one of $n$ possible solvers, where each solver $i$ has access to inventory of token $T_2$, at a price $p_i$ in token $T_1$. The user only trades with a solver if they offer a price improvement above the public market. We assume that $p_i$ are distributed i.i.d. according to a distribution with CDF $F(p)$ and density $f(p)$. This price could be either the price of their inventory or the price of an external market in which they have infinite or arbitrarily large liquidity relative to the user's demand. The user, in an attempt to elicit the best price for their tokens and improve upon the prices available from the public market, runs an auction (such as the Dutch auction found in UniswapX~\cite{Adams_2023}). We will examine several settings, in increasing order of complexity. 

\paragraph{Entry costs, costly effort, and congestion.}

Solvers face a sequence of choices in attempting to fill a user's order. First, they must choose whether or not to enter the auction in the first place. This choice is often driven by initial setup and infrastructure costs that are required to bid on and execute a user's orders. We call these \emph{entry costs}. These entry costs determine an equilibrium number $\kstar$ of the universe of $n$ solvers that will enter the auction. 
Upon making the choice to enter, the solvers then must decide how to bid. This choice, in our model, is driven by the fact that solvers expend costly effort to search for a price to win and satisfy a user's order, which modifies the distribution of prices to $F(p, e)$, where $e$ is the effort level chosen by the solver. In this case, solvers may try to simultaneously access a set of external markets, and therefore, effort may cause congestion. We seek to understand how the solvers' bidding behavior changes as a function of these two kinds of costs they face. In summary, we have the following settings and their associated results:
\begin{enumerate}
    \item \emph{No cost of entry, no effort costs}:  No cost of entry implies that the equilibrium number of entrants into the solver market satisfies $\kstar = n$. Classical revenue equivalence~\cite{krishna2009auction} applies and user welfare is maximized. 
    \item \emph{Costly entry}:  In equilibrium, the intent market has an asymptotically negligible number of entrants $k^*$ (that is, $k^*/n \rightarrow 0$ as $n \rightarrow \infty$) for reasonable distributions like the exponential and uniform distributions. For the Pareto distribution, however, in anticipation of large prices, we may have $k^* = n$ entrants. 
    \item \emph{Costly effort}: We demonstrate cases welfare is reduced due to effort causing congestion. This indicates that a planner may in fact want to restrict entry into such a market. 
\end{enumerate}

These results demonstrate that entry and effort costs may lead to oligopoly, defined as the combination of reduced competition ($\kstar = o(n)$) and reduced welfare or auction revenue. Critically, in the case of effort costs, a social planner may wish to limit entry, as a closed or permissioned solver market may achieve better outcomes from the perspective of a welfare-maximizing planner. 
Understanding when this occurs in practice, and how such permissioned markets can be implemented is an important design consideration that we leave for future work.

\subsection{Dutch auctions for intents} 

The Dutch action has been implemented by by various protocols, including UniswapX~\cite{Adams_2023}, to execute a single user's swap intent in a market with multiple solvers. It is well-known that the Dutch auction is strategically equivalent to a sealed-bid first price auction~\cite{krishna2009auction}. Therefore, we analyze the solvers' strategies in the setting of a first price auction, where the highest bidding solver wins at the highest price. Each solver submits a price $\tilde{p}_i \leq p_i$, which is their bid. According to the rules of the first price auction, the highest bid gets filled as long as it is better than $p^*$. The net profit of a solver from being filled therefore becomes:
\[ p_i  - \tilde{p}_i.\]
If $k$ of the universe of $n$ possible solvers are present in the auction, revenue equivalence yields that
\[
(p_i - \tilde{p}_i) F^{k-1} (p_i) = \int_{p^*}^{p_i} F^{k-1}(x) dx,
\]
or, equivalently, 
\[
\tilde{p}_i = (p_i ) - \frac{ \int_{p^*}^{p_i} F^{k-1}(x) dx}{F^{k-1} (p_i)}.
\]
This formula has a very simple interpretation: in a first price auction, in the absence of entry or congestion costs, each solver shades their bid according to the competition they face in the auction.

\paragraph{Interim profits.} In this case, the interim expected profit of a solver with price $p_i$ in a setting with $k$ other solvers can be written as:
\[
S(p_i, k) = \int_{p^*}^{p_i} F^{k}(x) dx.
\] 
We can therefore write the ex-ante expected profit of a solver as
\[
S(k) = \int_{0}^{\infty} S(p, k) f(p) dp.
\]
Note that:
\begin{align*}
    S(k) &= \int_{0}^{\infty} S(p, k) f(p) dp \\
              &= - \int_{0}^{\infty} S(p, k) \frac{d (1-F(p))}{dp} dp \\
              &= - \left( S(p, k) (1-F(p))\big\vert_{p=0}^{p =\infty}  -  \int_{0}^{\infty} S'(p, k) (1-F(p)) dp \right) \\
              &= \int_{p^*}^{\bar{p}} F^{k}(p) (1-F(p)) dp.
\end{align*}
Therefore, by observation, $S$ is decreasing and satisfies increasing differences in $k$. Note that $S: \mathbf{N} \rightarrow \mathbf{R}$ can be extended to the positive reals using~\eg, linear interpolation between successive integers. The resulting function is convex, and we abuse notation to consider this extended function in the sequel.

\paragraph{Exponential distribution.}
First suppose that that $F$ is the exponential distribution with rate $\lambda$ and
$p^*=0$. Then,
\begin{align*}
    S(k) &= \int_0^\infty (1- \exp(-\lambda p))^k \exp(-\lambda p) dp \\ & = \frac{1}{\lambda} \int_0^1 x^k dx  && \\
    & = \frac{1}{(k+1) \lambda},
\end{align*}
where the second equality is by substituting $1- \exp (-\lambda p) = x$.

\paragraph{Uniform distribution.}
Now suppose that $F$ is the uniform distribution on [0,1] and $p^*=0$. Then,
\[S(k) = \int_0^1 p^k (1-p) dp, \]
which in turn implies that 
\[S(k) = \frac{1}{(k+1)(k+2)}.\]

\paragraph{Pareto distribution.} 
Finally, suppose that $F$ is the generalized Pareto distribution with CDF $F(x) = 1 - \left[1 + \left(\frac{x-\mu}{\sigma}\right)^{1/\gamma}\right]^{-\alpha}$, where $x > \mu$, and $\sigma, \gamma, \alpha > 0$, and that $p^* = 0$. We consider this distribution in the regime $\gamma / \alpha > 1$, for which its mean is infinity. In this case, we have that
\begin{align*}
    S(0) & = \int_{0}^{\infty} 1-F(p) dp \\ & =\Expect_{p \sim f}[p] \\ & = \infty.
\end{align*}
We will use this fact shortly to show that in this case, the market admits all $n$ potential solvers as entrants.

\subsection{Costly entry}\label{sec:costly-entry-no-effort}
Now, we assume that prior to bidding in the auction, every solver must pay an entry cost of $c_i$ if they choose to bid. These costs correspond to the setup and infrastructure costs considered previously. For now, we consider the case that the solver does not know how many other solvers have entered or their oracle price; they choose to enter only based on their own price and cost. Costs are nonnegative and distributed according to a given distribution $F_C$. In the subsequent, we will always assume that $F_C$ has a density $f_C$ for whom $f_C(0)$ is finite and nonzero. Under these entry costs, we will show that expected number of entrants in the auction, $\kstar$, is small relative to the universe of possible entrants, $n$. 

\paragraph{Market size as a function of costs.}
The above setup allows us to study entry into the market by providing a threshold cost $\bar{c}$ such that it is not profitable in expectation for a solver to enter the auction if their cost exceeds the threshold.
The maximum cost solver who will choose to enter this market will be one with cost $\bar{c}$ where $\bar{c}$ solves:
\begin{align}
\sum_{k=0}^{n} \left({n \choose k} F_C^{k}(\bar{c}) (1- F_C(\bar{c}))^{n-k}\right) S(k) = \bar{c}.
\label{eq:threshold}
\end{align}
Recall that solvers enter the auction if their cost is less than the threshold cost $\bar{c}$. This happens with cumulative probability $F_C(\bar{c})$. Therefore, once the threshold $\bar{c}$ has been found, the expected number of entrants into the solver market can be computed as:
\begin{align*}
    k^* = n F_C(\bar{c}).
\end{align*}
We now show how $k^*$ depends on $n$ for various realistic distributions $F$.

\paragraph{Exponential distribution.} For the case of exponentially distributed prices with rate $\lambda$, we have that the left hand side of the above equation \eqref{eq:threshold} can be written as:
\begin{align*}
    \sum_{k=0}^{n} \left({n \choose k} F_C^{k}(\bar{c}) (1- F_C(\bar{c}))^{n-k}\right) S(k)    = \frac{1}{(n+1)\lambda F_C(\bar{c})} (1- (1-F_C(\bar{c}))^{n+1}).
\end{align*}
This equation is derived in Appendix \ref{app:exponential-distribution}. So, for exponentially distributed prices we must have that
\begin{align*}
   &\frac{1}{(n+1)\lambda F_C(\bar{c})} (1- (1-F_C(\bar{c}))^{n+1}) = \bar{c},
\end{align*}
which implies 
\begin{align*}
 \frac{1}{(n+1)\lambda} = \frac{\bar{c}F_C(\bar{c})}{(1- (1-F_C(\bar{c}))^{n+1})}.
\end{align*}
From this equation, we straightforwardly have that $\bar{c}$ must be decreasing to 0 as $n \nearrow \infty$. This is because $\bar{c}$ asymptoting to a constant would result in the left hand side going to 0 while the right hand side stays constant. Now suppose that $\bar{c} = g(n)$ for some decreasing function $g$ and take $n$ large so that we can do Taylor approximations. Then, the above equation becomes
\begin{align*}
   &\frac{1}{(n+1)\lambda}   = \frac{(g(n))^2 f_C(0) }{(1- (1-f_C(0)g(n) )^{n+1})}.
\end{align*}
This implies that $g(n) = O(\frac{1}{\sqrt{n}})$ since for this rate of convergence the denominator of the right hand side converges to $1$ because $\left(1- f_C(0) g(n)\right)^{n+1} \rightarrow 0$ and the numerator matches the left hand side. Recall that the expected number of entrants into the solver market, $k^*$, equals $n F_C(\bar{c})$. Therefore, for large $n$, we have $k^* = O(\sqrt{n})$ for exponentially distributed values.

\paragraph{Uniform distribution.}
Now, for the case of prices uniformly distributed between $[0,1]$, we have that the left hand side of equation \eqref{eq:threshold} is:

\begin{align*}
        & \sum_{k=0}^{n} \left({n \choose k} F_C^{k}(\bar{c}) (1- F_C(\bar{c}))^{n-k}\right) S(k) \\ & = \frac{1}{(n+1)(n+2) F_C^2(\bar{c})} (1- (1-F_C(\bar{c}))^{n+2} - (n+2) F_c(\bar{c}) (1-F_C(\bar{c}))^{n+1})
\end{align*}
This equation is derived in Appendix \ref{app:uniform-distribution}. Therefore, for uniformly distributed prices, we must have that
\begin{align*}
    & \frac{1}{(n+1)(n+2) F_C^2(\bar{c})} (1- (1-F_C(\bar{c}))^{n+2} - (n+2) F_c(\bar{c}) (1-F_C(\bar{c}))^{n+1}) = \bar{c}, 
\end{align*}
which implies
\begin{align*}
   \frac{1}{(n+1)(n+2)} = \frac{\bar{c}F_C^2(\bar{c})}{(1- (1-F_C(\bar{c}))^{n+2} - (n+2) F_c(\bar{c}) (1-F_C(\bar{c}))^{n+1})}.
\end{align*}
Similar arguments as in the case of exponentially distributed prices now tell us that $\bar{c} = O(n^{-\frac{2}{3}})$ and therefore the expected number of entrants is $k^* = O(n^{1/3})$.

\paragraph{Pareto distribution.} Finally, when $F$ is the generalized Pareto distribution (in the  regime where $\gamma / \alpha > 1$), recall that this distribution has infinite mean, which means in particular that $S(0)$, the expected interim profits with no other participating solvers, is infinity. This immediately allows us to conclude that $\bar{c}$ in this case is also infinity, as we have: 
\begin{align*}
    (1- F_C(\bar{c}))^n S(0) \leq \sum_{k=0}^{n} \left({n \choose k} F_C^{k}(\bar{c}) (1- F_C(\bar{c}))^{n-k}\right) S(k) = \bar{c}.
\end{align*}
Because $S(0) = \infty$, the only $\bar{c}$ that makes the above inequality hold is $\bar{c} = \infty$. Therefore, because $F_C(\bar{c}) \rightarrow 1$ as $\bar{c} \rightarrow \infty$, we have that the total number of entrants into the auction is $k^* = n F_C(\bar{c}) = n$. 

\paragraph{Connections to extreme spacings.} The quantity $S(k)$, for a given distribution $F$, is intimately related to the expected difference between the highest and second highest price in a sample of $k+1$ prices drawn i.i.d. from $F$. This gap is also known as the \emph{extreme spacing}~\cite{mudholkar2009extremes}. For a distribution with CDF $F$ and PDF $f$, define for any $k \in \naturals$ the following quantity:  
\[
ES(k)  \equiv \Expect_{p_i \sim f}[p_{k:k} - p_{(k-1):k}],
\]
where $p_{k:k}$ and $p_{k-1:k}$ denote the largest and second-largest of $k$ i.i.d. draws according to $f$. We show in Appendix \ref{app:extreme-spacing} that if $p^* = 0$, then for any distribution, we have that
\begin{align*}
    (k+1) S(k) = ES(k+1).
\end{align*} 
For the generalized Pareto distribution, it was shown in ~\cite[Theorem 6.1]{mudholkar2009extremes} that $ES(k) = O(k^{\gamma/\alpha})$, which allows us to conclude $S(k) = O(k^{\gamma/\alpha - 1})$. Such bounds may allow one to derive rates for entry into the auction for a wider class of price distributions, which we leave for future work. 

\begin{figure}[h]
\begin{center}
    \includegraphics[scale = 0.45]{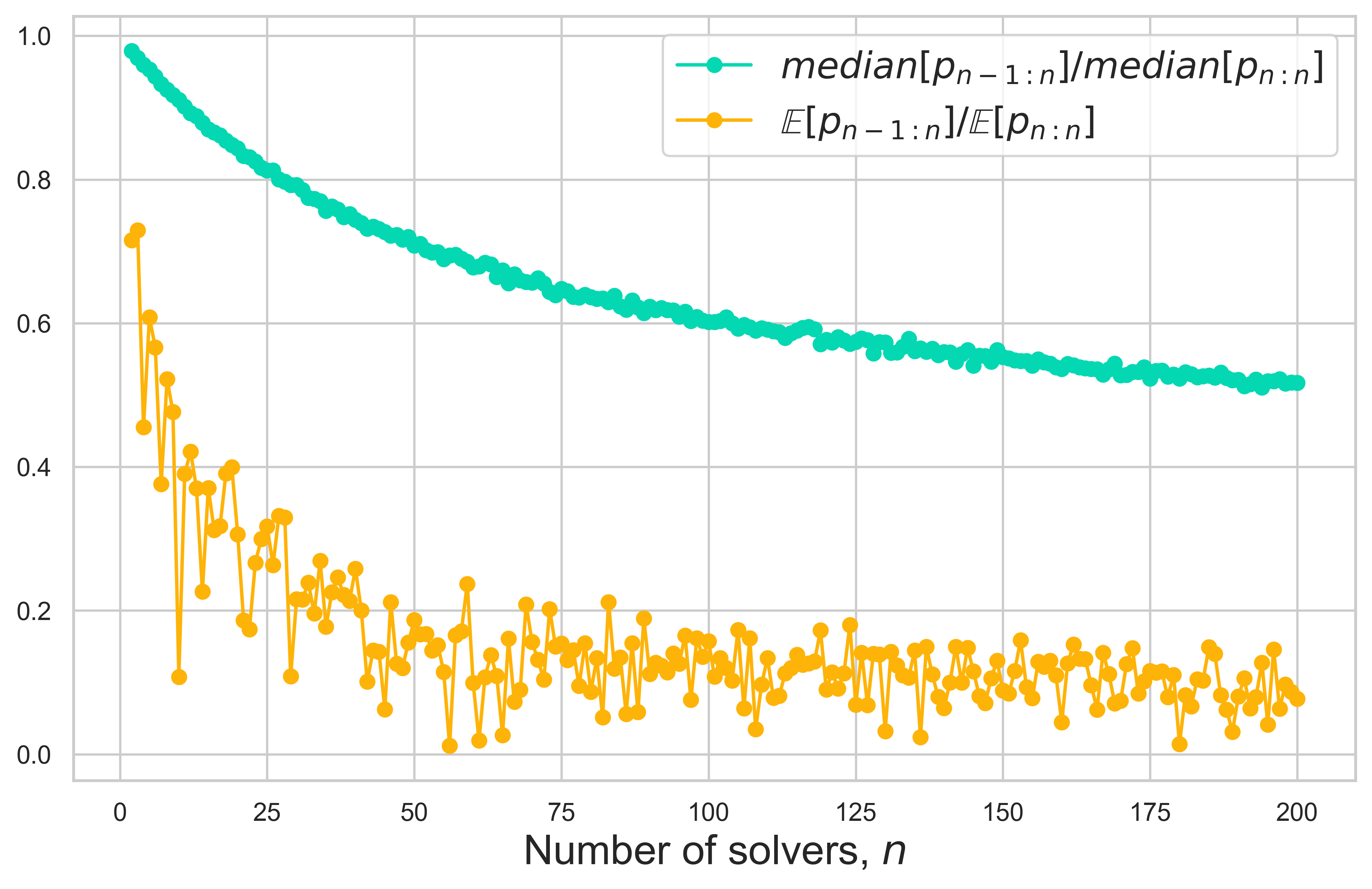}
    \caption{Ratio of the expected revenue to the expected highest price, $\Expect[p_{n-1:n}]/\Expect[p_{n:n}]$, for the Pareto distribution, versus the number of solvers, $n$. The ratio of the medians of $p_{n-1:n}$ and $p_{n:n}$ is plotted for reference.}
    \label{fig:pareto}
\end{center}

\end{figure}

\paragraph{Discussion and numerical results.} Our first set of results on costly entry demonstrate several interesting features of intent markets. For both the exponential and uniform distributions, entry costs can cause a much smaller fraction of potential solvers to enter the auction than would otherwise enter in the absence of costs. For reference, when 
$n = 100$, one may have as few as $k^* = 10$ entrants in the case of exponentially distributed prices, and as few as $k^* = 5$ entrants in the case of uniformly distributed prices. These results have a natural economic interpretation: solvers decide to enter the auction depending on whether they believe there is an opportunity to receive net expected profit that exceeds their threshold costs. In the case of the exponential distribution, the probability that a solver receives a price that is much better than another solver decays exponentially. The situation is even worse in the case of the uniform distribution, where there is an upper bound on the best price available to the solvers. 

This can be remedied in the case of the Pareto distribution, where even being a single solver in the market yields infinite expected profits. Therefore, in anticipation of these large profits, we can expect $n$ entrants into the market. We note, however, that even in this case, this does not mean that the user is able to realize a large price, as there may be significant bid shading in the auction. We demonstrate this effect in Figure \ref{fig:pareto}, where we plot the ratio of the expected second highest price (equivalent to the expected revenue of the auction, which in turn is the expected price the user receives) to the expected highest price, $\Expect[p_{n-1:n}]/\Expect[p_{n:n}]$, for $n$ i.i.d. draws of prices $p_i$ from the Pareto distribution. This ratio tells us how much the price the user can expect to receive from a solver auction with $n$ participants falls short of the theoretical highest price that could be achieved by any solver. The ratio of the medians of $p_{n-1:n}$ and $p_{n:n}$ are also plotted for reference. 

Our simulation is done by setting $\sigma = 100$, $\alpha = 0.95$, $\gamma = 1$, $\mu = 0$,  which corresponds to a standard Pareto distribution with CDF $F(x) = 1 - \left( \frac{x_m}{x} \right)^{\alpha}$, with scale parameter $x_m = 100$ and shape parameter $\alpha = 0.95$. For each $n$, we compute the expectations of $p_{n-1:n}$ and $p_{n:n}$ by averaging over $10000$ trials. Code for this simulation can be found at
\begin{center}
    \url{https://github.com/kkulk/intent}.
\end{center}
 We see that as $n$ gets large, the ratio $\Expect[p_{n-1:n}]/\Expect[p_{n:n}]$ approaches zero. This means that even with a large number of solvers present in the auction, the user is able to realize a price that is only a small fraction of the theoretical highest price available to any of the solvers. This suggests that intent markets in assets with heavy tails may fail to realize user welfare that is much greater than that afforded by public mechanisms such as CFMMs. 

\subsection{Costly effort}
So far, we have considered markets where costs of entry were fixed for each potential entrant, independent of the number of entrants, and similarly the distribution of prices for each entrant is exogenously given. 
Now, instead, suppose that entrants do not simply receive a price upon entry but have to invest costly effort.
This can be viewed as investment into infrastructure or technology in order to be able to participate in an auction---an example: a high-frequency trader who has to spend millions of dollars to achieve a sufficiently low message latency to successfully participate in an auction. Further, this can also correspond to the costs of sourcing liquidity for assets that the solver does not initially hold. As we will see, if these effort costs cause congestion (as they might in low-liquidity assets), user welfare may be harmed.

\paragraph{Effort-dependent prices and costs.}
Formally, suppose now that $k^*$ entrants have entered the auction (perhaps by the process of costly entry introduced in the previous section). Each entrant $i$ simultaneously chooses an investment level $e_i \in \reals_+$.
This investment results in the entrant drawing a price $p_i \sim F(p,e_i)$ for a CDF $F$ dependent on the amount of investment.
We make the assumption that investments improve values in the sense of first-order stochastic dominance: $F(p,e) \leq F(p,e')$ for any $p$ and $e > e'$.

Effort is both costly and congestive. We model the cost of taking effort $e$ when the total number of entrants is $\kstar$ by $c( \kstar,e)$ where $c$ is increasing and convex in both arguments. A special case of interest is when $c(\cdot)$ is constant in the first argument, so the cost of effort is not affected by other other entrants, \ie,~there is no congestion cost. Note that this setup is a symmetric game between the $\kstar$ solvers that enter the auction and thus admits a symmetric equilibrium, which we search for next.  

\paragraph{Equilibrium effort.}
\newcommand{\estar}{e^\star}
We now solve for the equilibrium investment $\estar$ by each entrant in this model.
To do so, we pick a solver $i$ and assume that all $\kstar-1$ competing solvers play the equilibrium strategy of drawing $p \sim F(x, \estar)$.
Given this strategy, the expected revenue of solver $i$ with with price $p$ is:
\[
S(p,\kstar,\estar) := \int_0^p F(x,\estar)^{\kstar-1} dx
\]
This expression corresponds to the probability that solver $i$'s value is higher than the remaining $\kstar-1$ solvers. Therefore, given that all the other entrants invest $\estar$, entrant $i$ must pick their level of investment $e_i$ to maximize their expected profit: 
\[ 
e_i = \argmax_{e\in\mathbf{R}_+} \int S(p,\kstar,\estar) f(p,e) dp  - c(\kstar,e).
\]
Assuming that $\estar >0$, we must also have $e_i$ satisfy a first order condition at $e = \estar$.
This condition, assuming that the densities and costs are differentiable, is:
\begin{equation}\label{eq:foc}
\left(\int S(p,\kstar,\estar) \frac{\partial f(p,e)}{\partial e} dp - \frac{c(\kstar,e)}{\partial k}\right) \Bigg\rvert_{e = \estar} =0.
\end{equation}
This equation implicitly describes $\estar$ as a function of $\kstar$, which allows us to measure congestion and effort's impacts on net revenue.

\paragraph{Properties of the equilibrium.}

We now consider the following specific example: $F(p,e) = p^e$ for $p \in [0,1]$. Note that therefore $f(p,e) = e p^{e-1}$.
Informally, an investment of $e$ results in $e$ draws from a uniform distribution, with the entrant's value being the maximum of these.
We assume from our results in~\S\ref{sec:costly-entry-no-effort} that $\kstar = o(n)$ solvers have entered the auction. 

Further, suppose $c(\kstar, e) = \alpha(\kstar) \frac{e^2}{2}$, where $\alpha(\cdot)$ is an increasing function which parametrizes the `congestiveness' of the underlying market as a function of the number of entrants in the market.
This functional form matches one's intuition for inventory cost: if all the entrants have to carry inventory in an intent market, then the existence of other entrants drives up the costs of sourcing and carrying the inventory.  
In this case we have
\[
S(p,\kstar,\estar)= \frac{p^{(\kstar-1)\estar + 1}}{(\kstar-1)\estar + 1}
\]
Substituting this into our first order condition~\eqref{eq:foc} we therefore have
\[
\int_0^1 \frac{p^{(\kstar-1)\estar + 1}}{(\kstar-1)\estar + 1} (p^{\estar-1} +\estar p^{\estar-1} \ln p) dp - \alpha(\kstar)\estar =0
\]
Solving this integral yields the following equation:
\begin{align*}
    \frac{1}{(1+ \estar \kstar)^2} - \alpha(\kstar) \estar =0 \hspace{0.2cm}\implies& \hspace{0.2cm} \alpha(\kstar) \estar (1+ \estar \kstar)^2 =1.
\end{align*}
Since $\alpha(\cdot)$ is nondecreasing, we must have that the solution $\estar$ is decreasing in the number of entrants $\kstar$. Note that we can write the left hand side as
\begin{align*}
\frac{\alpha(\kstar)}{\kstar} \left(\estar \kstar (1+ \estar \kstar)^2 \right) =1
\end{align*}
Observe that the function $\estar \kstar (1+ \estar \kstar)^2$ is increasing in $\estar \kstar$. 
This implies that, viewing $\estar$ as an implicit function of $\kstar$, if $\alpha(\kstar)$ is linear (\ie~$\alpha(\kstar) = \Theta(\kstar)$), we have that $\estar \kstar$ is constant in $\kstar$ or $\estar = c/\kstar$ for some constant c.
Conversely, if $\alpha(\kstar)$ grows faster (slower) than linear, (\ie,~$\alpha(\kstar)/ \kstar$ is increasing (decreasing) in $\kstar$), then we have that $\estar \kstar$ is decreasing (increasing) in $\kstar.$ 
Since $\kstar = \omega(1)$, this implies that if $\alpha(\kstar) = \Omega(\kstar)$ then $\estar = o(1)$ in $\kstar$ (and hence, $n$).
In other words, a larger universe of searchers leads to vanishing expected equilibrium effort as $\frac{\estar}{\kstar} \rightarrow 0$.

\paragraph{Welfare as a function of effort.}
We now analyze the welfare of the user as a function of effort $\estar$ and equilibrium participation $\kstar$.
Note that user welfare is maximized when the expected revenue of the auction is maximized, since that is the (expected) price that the user receives.
By revenue equivalence, the expected revenue equals the revenue of a second-price auction, so we can write the revenue as:
\begin{align}
    \mathsf{Rev}(\estar, \kstar) = &\int_0^1 p \kstar (\kstar- 1) (1-F(p,\estar)) f(p,\estar) (F(p,\estar))^{\kstar - 2} dp \nonumber\\
    =& \int_0^1 p \kstar (\kstar- 1) (1-p^{\estar}) (\estar p^{\estar-1}) (p^{\estar})^{\kstar - 2} dp \nonumber\\
    =& \frac{{\estar}^2 (\kstar-1)(\kstar)}{(1+\estar(\kstar-1))(1+\estar \kstar) } \nonumber\\
    =& \frac{{\estar} (\kstar-1)}{(1+\estar(\kstar-1))}\times\frac{{\estar} \kstar}{(1+\estar \kstar) }\label{eq:rev-kstar}
\end{align}
We make two claims about this formula:
\begin{enumerate}
    \item If $\alpha(\kstar) = o(\kstar)$, then $\mathsf{Rev}(\estar, \kstar)$ increases in $\kstar$
    \item If $\alpha(\kstar) = \omega(\kstar)$, then $\mathsf{Rev}(\estar, \kstar)$ decreases in $\kstar$
\end{enumerate}
The first claim states that if the congestion cost is sublinear in $\kstar$, then the planner has no reason to restrict entry into the auction, since every new entrant increases their revenue.
On the other hand, the second claim states that if the congestion cost is superlinear in $\kstar$, then the user would prefer to \emph{restrict entry}, \ie~a limited oligopoly makes the user better off.

We show that the two claims are correct using elementary arguments.
The first claim is equivalent to the statement: if $\alpha(\kstar) = o(\kstar)$, then $\frac{\partial \mathsf{Rev}(\estar, \kstar)}{\partial\kstar} > 0$.
To show this, note that if $\alpha(\kstar) = o(\kstar)$, then $\estar \kstar$ is increasing in $\kstar$ and so is $\estar(\kstar - 1)$.
Therefore both terms in~\eqref{eq:rev-kstar} are increasing in $\kstar$.
Similarly, the second claim states that if $\alpha(\kstar) = \omega(\kstar)$ then $\frac{\partial \mathsf{Rev}(\estar, \kstar)}{\partial\kstar} < 0$.
Since $\alpha(\kstar) = \omega(\kstar)$ implies that $\estar \kstar$ is decreasing in $\kstar$ and $\estar(\kstar -1)$ is decreasing in $\kstar$.
This implies $\frac{\partial \mathsf{Rev}(\estar, \kstar)}{\partial\kstar} < 0$.

\section{An optimization approach}
Here, we present an alternative, deterministic model of the intent market, which
follows from an optimization framework. Instead of considering prices for each
solver, drawn from some distribution, we equip each solver with a utility and
cost function. This approach has a number of advantages. First, these functions
can be easily parameterized, and, as a result, the approach lends itself to
empirical studies and model fitting with real-world data. Second, modeling with
utility functions may be more appropriate in some scenarios, particularly when
there is a lot of heterogeneity among the solvers. Finally, this approach
immediately suggests a Dutch auction-like mechanism, analogous to that
considered in the probabilistic model. We also show how, under certain
assumptions, the deterministic framework implies similar results as those in the
previous section, such as those involving the impact of congestion.

In particular, we turn to convex optimization for our deterministic approach,
which is similar in spirit to the linear programs presented
in~\cite{kalagnanam2004auctions} in the context of auctions. We outline how
social welfare maximizing allocations may be reached via a primal-dual
mechanism, which can be interpreted as a Dutch auction. As in the previous
section, we limit our focus to two-asset trades. In the general multiple asset
case, one must use a more complicated mechanism, as asset prices may be
negatively correlated. We leave this case to future work.

\subsection{Social welfare}

\paragraph{Solvers.}
Each solver $k = 1, \dots, n$ is willing to provide, for given amount $x$ of
token $T_1$, an amount $\alpha_k x$ of $T_2$. To do this, the solver incurs a
cost $c_k(x)$ and gains utility $u_k(x)$, where the function $u_k$ may, for
example, reflect the solver's relative preference for holding $T_1$ versus
$T_2$. We restrict the cost function $c_k: \reals_+ \to \reals_+ \cup
\{\infty\}$ to be convex, the utility function $u_k: \reals_+ \to \reals_+$ to
be concave, and both to be defined over the nonnegatives---negative trades are
not allowed. Infinite values are used to encode constraints; a trade $x$ such
that $c_k(x) = \infty$ is unacceptable (\eg, it is outside solver $k$'s budget).
We assume that these functions are closed and proper. Thus, the net utility for
solver $k$ to tender $\alpha_k x_k$ units of $T_2$ to the user in exchange for $x$
units of $T_1$ is given by
\[
U_k(x_k) = -\alpha_k x_k - c_k(x_k) + u_k(x_k),
\]
where this utility is measured in terms of token $T_2$, which we take as the
numeraire. This utility is concave by construction.

\paragraph{User.}
We consider a user who wishes to trade $\delta$ units of token $T_1$ for a
maximum amount of token $T_2$, using a combination of these solvers and an
on-chain CFMMs. The user can find the optimal trade with the network of CFMMs by
solving the \emph{optimal routing
problem}~\cite{angeris2022optimal,diamandis2023efficient}. Because this problem
is a convex optimization problem with a concave objective, we can partially
maximize~\cite[\S3.2.5]{boyd2004convex} to define an aggregate forward exchange
function $G: \reals_+ \to \reals_+$, which maps an input of token $T_1$ to the
maximum output of token $T_2$ using all of the on-chain CFMMs. Note that this
function is concave and nondecreasing. The user's utility is then given by
\[
    U_p(x) = G(\delta - {\textstyle\sum_{k=1}^K x_k}) + \sum_{k=1}^K \alpha_k x_k,
\]
which is concave and nondecreasing.

\paragraph{Social welfare maximization.}
Putting these utility functions together, we can define the problem of
maximizing social welfare:
\begin{equation}
\label{eq:opt-social-welfare}
\begin{aligned}
    &\text{maximize} && G(\delta - y) 
    + \sum_{k=1}^n u(x_k) - c_k(x_k),\\
    &\text{subject to} && y = \sum_{k=1}^n x_k.
\end{aligned}
\end{equation}
with variables $y \in \reals$ and $x_k \in \reals$ for $k = 1, \dots, n$. Note
that the transfer from each solver $k$ to the user, given by $\alpha_k x_k$, has
canceled out in the objective.

\subsection{Dutch auction}
The Lagrangian of the social welfare maximization problem~\eqref{eq:opt-social-welfare} is
\[
L(x, \tilde x, \nu) = 
G(\delta - y) + \nu y + \sum_{k=1}^n u_k(x_k) - c_k(x_k) - \nu x_k.
\]
The dual function is then
\[
\begin{aligned}
h(\nu) &= 
\sup_{y \ge 0}\left\{G(\delta - y) + \nu y \right\} 
+ \sum_{k=1}^n \sup_{x_k \ge 0}\left\{u_k(x_k) - c_k(x_k)  - \nu x_k \right\}.
\end{aligned}
\]
Referring back to the utilities, we can interpret these terms simply: they are
exactly the utility for the user and agent respectively when the universal price
of the asset is $\nu$. Since the function $G$ is nondecreasing, we can conclude
that the price $\nu \ge 0$. Otherwise if $\nu < 0$, letting $y = -t$ we have 
\[
    G(\delta + t) - \nu t \ge G(\delta) + \lvert{\nu}\rvert t,  
\]
and the right hand side goes to infinity as $t \to \infty$ (equivalently, as $y
\to -\infty$).

\paragraph{Optimality conditions.}
From the Lagrangian, at optimality we have primal feasibility ($y^\star =
\sum_{k=1}^n x_k^\star$) and the dual conditions
\[
\begin{aligned}
     g(\delta - y^\star) &= \nu^\star  \\
     u_k'(x_k^\star) - c_k'(x_k^\star) &=  \nu^\star \quad \text{for} ~ k = 1, \dots, n
\end{aligned}
\]
(When $u_k$ or $c_k$ is not differentiable for some $k$, a similar argument
holds using subgradient calculus.) Putting these together, we have that 
\[
u_k'(x_k^\star) - c'_k(x_k^\star) = g(\delta - \textstyle \sum_{k=1}^n x_k).
\]
In other words, the marginal cost of getting asset 2 from the public CFMMs
should exactly equal each participating solver's net marginal utility for
providing it. These optimality conditions allow us to also characterize when a
solver will not participate for a given price $\nu$. Any solution to the solver
subproblem, denoted by $\tilde x_k$, which lies on the interior of the domain
($\tilde x_k > 0$) must satisfy
\[
u'(\tilde x_k) - c'(\tilde x_k) =  \nu.
\]
Thus, if $\nu$ is set too high, the solver will simply return the zero trade:
$\tilde x_k = 0$. Furthermore, this condition implies that when the costs to the
solver are zero, it will report its true marginal utility (\ie, its true price),
corroborating the results in \S\ref{sec:costly-entry-no-effort}.

\paragraph{A dual approach.}
We can solve this social welfare problem with a mechanism akin to a dutch
auction. As a consequence of strong duality, which holds under relatively weak 
conditions~\cite[\S5.2.3]{boyd2004convex}, we can solve the primal
problem~\eqref{eq:opt-social-welfare} by solving the dual problem. The dual
problem is
\[
\begin{aligned}
    &\text{minimize}    && h(\nu).
\end{aligned}
\]
Since this problem is convex and single dimensional, we can solve it with an
iterative method that can be interpreted as a dutch-auction-like mechanism. The
first derivative of $h$ is 
\[
    h'(\nu) = \tilde y - \sum_{k=1}^n \tilde x_k,
\]
where $\tilde y$ is the optimizer of $G(\delta - y) + \nu y$ and $\{\tilde
x_k\}$ are the optimizers of the solvers' subproblems, \ie, the amount they will
quote given a price $\nu$. Since this function is unimodal, we can optimize it
with a simple strategy: we start with a large $\nu$ and query each solver to get
$\{\tilde x_k\}$, which is exactly the solution to the solver
utility-maximization subproblem above. Then, we compute the gradient $h'(\nu)$.
If this gradient is greater than zero, we decrease $\nu$ and repeat. When the
gradient is sufficiently close to zero, we are done. In practice this looks like
a Dutch auction. Note that, at termination, the $\tilde y$ and $\{\tilde x_k\}$
are exactly the solutions to the original social welfare
problem~\eqref{eq:opt-social-welfare}.
The dual setup above can also be extended to the case of $n$ assets with
independent solver costs. However, the proposed mechanism does not extend
directly, as there is no longer a total ordering on prices. A coordinate descent
type mechanism may work in this multiple asset case.

\paragraph{Interpretation.}
Recall that $\alpha_k$ is the price a solver quotes to the user. It is clear
that for any rational solver
\[
    u_k(\delta) \ge \alpha_k \delta \implies u_k(\delta)/\delta \ge \alpha_K.
\]
In other words, $\alpha_k x_k$ is a linear underestimator of the solver's
concave utility function $u_k$. The mechanism above, then, elicits the solver's
true price for a given order size, and it terminates when the solver's marginal
price is equal to that of the CFMM. However, instead of proposing an order size
for each of the $n$ solvers, the mechanism instead can propose one price, which
it broadcasts to all solvers. The solvers respond to this price with a proposed
order.

\subsection{Extension: congestion}
From this optimization framework, we also see that positively correlated costs
among solvers will reduce the user's welfare. Now assume that the solver cost
functions are not independent. We model congestion by introducing a new cost
function $\tilde c_k:\reals^n_+ \to \reals_+ \cup \{\infty\}$ which is
increasing in all arguments and matches the old cost function $c_k$ when that
solver is the only one quoting an order. In other words, we must have that 
\[
    \nabla \tilde c(x_1, \dots, x_n) > 0
\]
and
\[
    \tilde c_k(0, \dots, x_k, \dots 0) = c_k(x_k)
\]
for any order $x_k$.
These conditions imply that marginal costs are higher in this regime:
\[
    \tilde c_k'(x_1, \dots, x_n) > \tilde c_k(0, \dots, x_k, \dots 0) = c_k'(x_k),
\]
where there is at least one solver $k' \neq k$ with $x_{k'} > 0$.
As a result, from the dual conditions, we have that 
\[
    \nu^\star = u_k'(x_k^\star) - \tilde c_k'(x_1^\star, \dots, x_K^\star) 
    \le u_k'(x_k^\star) - c_k'(x_k).
\]
Thus, the price $\nu^\star$ when congestion costs are present is less that the
price with independent costs. In other words, the user will receive reduced
output from the solvers if their costs are positively correlated. This result
mirrors that of the probabilistic model: under congestion costs, user welfare
decreases.

\section{Model Extensions}\label{sec:model-extensions}
As discussed in the introduction, there are both non-blockchain focused intent markets as well as more complex intent-like systems within the blockchain world.
We will briefly discuss the applicability of our model to those scenarios and extensions that can be used for analyzing these systems.

\paragraph{Stock order-by-order auctions.}
The US Securities and Exchange Commission (SEC) recently released a proposal for an order flow execution auction within the US stock market~\cite{sec-prop-605}.
One rationale for implementing such a system is to improve the fairness of `price improvement' that a retail stock trader receives~\cite{sec-prop-605}. 
The majority of retail, individual stock traders within the US use brokerages such as Robinhood or Vanguard to execute their trades.
These brokerages outsource the task of finding an optimal execution price for a user to wholesalers, who aggregate order flow from brokerages and institutions (such as endowments or pension funds) to execute trades.
Currently, there is no competitive market for brokerages to engage wholesalers---they instead construct private agreements which are known as payment for order flow~\cite{welch2022wisdom}.
These agreements have come under scrutiny by the US government, as the lack of a competitive market is thought to have led to market failures during the Gamestop panic of 2021~\cite{angel2021gamestonk}.

The SEC's proposal is to have wholesalers run an auction whose initial price is set at the price tendered by the wholesaler.
The wholesaler would run the auction for 100 milliseconds and any regulated entity, such as an accredited investor, could participate in the auction.
To win the auction, the wholesaler would have to participate explicitly, as their initial price does not serve as a reserve price. In contrast, the UniswapX Dutch auction guarantees the user at least the amount of output quoted publicly by a CFMM, even if the auction does not clear.

Due to latency constraints, it is assumed that the wholesaler would be running many of these auctions simultaneously. It is an interesting extension of our model to understand the degree to which auctions in multiple markets, where price movements may be correlated, result in improved prices for users.

\paragraph{Proposer-builder separation.}
One of the most common auctions settled on blockchains is the so-called `proposer-builder separation' (PBS) auction~\cite{pai2023structural, gupta2023centralizing, bahrani2024centralization}.
The seller in this auction is the proposer: an agent who has a temporary monopoly to propose a valid block within a blockchain.
The proposer runs an auction to solicit blocks from agents known as builders who source transactions and construct valid blocks.
A number of works have studied the centralizing effect within PBS auctions~\cite{bahrani2024centralization, gupta2023centralizing}.
However, none of the prior works analyzed the impact of congestion on auction performance.

PBS auctions fit into our framework as they are single item first price auctions.
Block builders, who source transactions to earn fees or rewards from block construction (\ie~ the aforementioned MEV ~\cite{kulkarni2022towards, bahrani2023transaction, kulkarni2023routing}), compete for order flow from exchanges and wallets that generate transactions.
Similar to PFOF, these builders often offer incentives to users to send their transactions to them, and may exert costly effort to acquire user order flow.
One notable difference is that our model assumes a uniform congestion cost across agents.
In the PBS auction, different agents will have varied costs and congestion is more heterogeneous across agents.
Modifying our model to include hetereogeneity between the agents is future work and can be used to analyze historical auction data to measure reductions in competition.

\section{Conclusion}
In this paper, we model the incentives that solvers face in intent-based markets.
Our first model is a probabilistic model of an auction under two complicating features: \textit{(i)} entry costs incurred by solvers while setting up infrastructure and \textit{(ii)} congestion costs faced by solvers while attempting to search for liquidity at desirable prices.
In our single intent model (single item auction model), we demonstrate that entry costs reduce the number of solvers that participate in the market in equilibrium.
If, furthermore, participants have to invest costly effort into improving their ability to find prices, and face congestion costs while doing so, we also find the existence of limited oligopoly.
These results demonstrate the counterintuitive fact that planners of intent markets may want to limit entry in order to maximize user welfare.

Our second model is an optimization-based deterministic model. This model enables the explicit analysis of the social welfare maximization problem faced by a planner who considers both the user's and the solvers' utilities. This approach has several benefits: \textit{(i)} it allows for fitting utility functions to data and \textit{(ii)} suggests a natural Dutch auction-like primal-dual mechanism to solve the welfare maximization problem.

We leave as future work a generalization of the single intent model to more complicated intents which may include transacting in multiple markets. Further, it is an interesting question whether numerical methods may be used to estimate whether solvers face congestion costs in these markets, and the degree to which users face welfare losses. Finally, in our modeling so far, motivated by practice, we have restricted attention to solvers competing in an auction to fulfill the user's intent, where the profits from the auction provide the incentives for solvers to enter and invest. However, certain kinds of congestive investment costs may be difficult to observe and even more challenging for users to directly contract \cite{grossman1992analysis}. Recent developments in the literature provide insight into how a principal can robustly incentivize effort by an agent (\eg, \cite{carroll2015robustness}). Another direction of future work, therefore, is whether such contracts can be implemented in intent markets.

Our work implies that designers and planners of intent markets need to thoroughly vet the various costs that solvers face. 
If the goal of an intent network is to maximize welfare while also allowing free entry into the system, designers should be careful to construct auctions such that bidders minimize contention in sourcing inventory.

\section*{Acknowledgments}
We would like to thank Malcolm Granville for valuable feedback on an earlier version of this paper. We would also like to thank Karthik Srinivasan and Barry Plunkett for helpful discussions. 
\printbibliography

@inproceedings{angeris2022optimal,
  title={Optimal routing for constant function market makers},
  author={Angeris, Guillermo and Evans, Alex and Chitra, Tarun and Boyd, Stephen},
  booktitle={Proceedings of the 23rd ACM Conference on Economics and Computation},
  pages={115--128},
  year={2022}
}

@article{diamandis2023efficient,
  title={An Efficient Algorithm for Optimal Routing Through Constant Function Market Makers},
  author={Diamandis, Theo and Resnick, Max and Chitra, Tarun and Angeris, Guillermo},
  journal={Financial Cryptography},
  year={2023}
}

@article{angeris2019analysis,
  title={An analysis of Uniswap markets},
  author={Angeris, Guillermo and Kao, Hsien-Tang and Chiang, Rei and Noyes, Charlie and Chitra, Tarun},
  journal={arXiv preprint arXiv:1911.03380},
  year={2019}
}

@inproceedings{angeris2020improved,
  title={Improved price oracles: Constant function market makers},
  author={Angeris, Guillermo and Chitra, Tarun},
  booktitle={Proceedings of the 2nd ACM Conference on Advances in Financial Technologies},
  pages={80--91},
  year={2020}
}

@inproceedings{milionis2022quantifying,
  title={Quantifying loss in automated market makers},
  author={Milionis, Jason and Moallemi, Ciamac C and Roughgarden, Tim and Zhang, Anthony Lee},
  booktitle={Proceedings of the 2022 ACM CCS Workshop on Decentralized Finance and Security},
  pages={71--74},
  year={2022}
}

@article{cartea2023predictable,
  title={Predictable losses of liquidity provision in constant function markets and concentrated liquidity markets},
  author={Cartea, {\'A}lvaro and Drissi, Fay{\c{c}}al and Monga, Marcello},
  journal={Applied Mathematical Finance},
  pages={1--25},
  year={2023},
  publisher={Taylor \& Francis}
}

@incollection{grossman1992analysis,
  title={An analysis of the principal-agent problem},
  author={Grossman, Sanford J and Hart, Oliver D},
  booktitle={Foundations of Insurance Economics: Readings in Economics and Finance},
  pages={302--340},
  year={1992},
  publisher={Springer}
}

@article{angeris2021replicating,
  title={Replicating monotonic payoffs without oracles},
  author={Angeris, Guillermo and Evans, Alex and Chitra, Tarun},
  journal={arXiv preprint arXiv:2111.13740},
  year={2021}
}

@article{lehar2023liquidity,
  title={Liquidity fragmentation on decentralized exchanges},
  author={Lehar, Alfred and Parlour, Christine and Zoican, Marius},
  journal={arXiv preprint arXiv:2307.13772},
  year={2023}
}

@article{adams2021uniswap,
  title={Uniswap v3 core},
  author={Adams, Hayden and Zinsmeister, Noah and Salem, Moody and Keefer, River and Robinson, Dan},
  journal={Tech. rep., Uniswap, Tech. Rep.},
  year={2021}
}

@misc{yin_2023, 
    title={An overview of Anoma’s architecture}, 
    url={https://medium.com/anomanetwork/an-overview-of-anoma-s-architecture-26b72e8c9be5}, 
    journal={Medium}, 
    publisher={Anoma | Intent-centric Architecture}, 
    author={Yin, Awa Sun}, 
    year={2023}, 
    month={9}
}

@article{kulkarni2022towards,
  title={Towards a theory of maximal extractable value i: Constant function market makers},
  author={Kulkarni, Kshitij and Diamandis, Theo and Chitra, Tarun},
  journal={arXiv preprint arXiv:2207.11835},
  year={2022}
}

@inproceedings{xavier2023credible,
  title={Credible Decentralized Exchange Design via Verifiable Sequencing Rules},
  author={Xavier Ferreira, Matheus Venturyne and Parkes, David C},
  booktitle={Proceedings of the 55th Annual ACM Symposium on Theory of Computing},
  pages={723--736},
  year={2023}
}

@inproceedings{daian2020flash,
  title={Flash boys 2.0: Frontrunning in decentralized exchanges, miner extractable value, and consensus instability},
  author={Daian, Philip and Goldfeder, Steven and Kell, Tyler and Li, Yunqi and Zhao, Xueyuan and Bentov, Iddo and Breidenbach, Lorenz and Juels, Ari},
  booktitle={2020 IEEE Symposium on Security and Privacy (SP)},
  pages={910--927},
  year={2020},
  organization={IEEE}
}

@misc{Adams_2023, 
   title={Introducing the UniswapX protocol}, 
   url={https://blog.uniswap.org/uniswapx-protocol}, 
   journal={Uniswap Protocol}, 
   author={Adams, Hayden}, 
   year={2023}, 
   month={7}
}

@misc{Protocol_2023, 
   title={Introducing the Programmatic Order Framework, from CoW Protocol}, 
   howpublished={\url{https://blog.cow.fi/introducing-the-programmatic-order-framework-from-cow-protocol-088a14cb0375}}, 
   journal={Medium}, 
   publisher={Medium}, author={Protocol, CoW}, year={2023}, month={11}
}

@misc{CoW_2023, 
    title={CoW Protocol --- Dune}, 
    howpublished={\url{https://dune.com/cowprotocol}}, 
    journal={CoW Protocol Statistics}, 
    publisher={Dune Analytics}, 
    author={CoW, Protocol}, 
    year={2023}, 
    month={12}
}

@misc{UniswapX-dune, 
    title={UniswapX --- Dune}, 
    howpublished={\url{https://dune.com/phu/uniswapx}}, 
    journal={UniswapX Statistics}, 
    publisher={Dune Analytics}, 
    author={\@phu}, 
    year={2024}, 
    month={01}
}

@article{mudholkar2009extremes,
  title={Extremes, extreme spacings and tail lengths: an investigation for some important distributions},
  author={Mudholkar, Govind S and Chaubey, Yogendra P and Tian, Lili},
  journal={Calcutta Statistical Association Bulletin},
  volume={61},
  number={1-4},
  pages={243--266},
  year={2009},
  publisher={SAGE Publications Sage India: New Delhi, India}
}

@article{bahrani2024centralization,
  title={Centralization in Block Building and Proposer-Builder Separation},
  author={Bahrani, Maryam and Garimidi, Pranav and Roughgarden, Tim},
  journal={arXiv preprint arXiv:2401.12120},
  year={2024}
}

@misc{Sengupta_2023, 
  title={Oracles and the New Frontier for Application-owned Orderflow auctions}, 
  url={https://multicoin.capital/2023/12/14/oracles-and-the-new-frontier-for-application-owned-orderflow-auctions/}, 
  journal={Multicoin Capital}, 
  publisher={Multicoin Capital}, 
  author={Sengupta, Shayon}, 
  year={2023},
  month={Dec}
}

@misc{Lambur_2024, 
  title={Announcing oval: Earn protocol revenue by capturing Oracle mev}, 
  url={https://medium.com/uma-project/announcing-oval-earn-protocol-revenue-by-capturing-oracle-mev-877192c51fe2}, 
  journal={Medium}, 
  publisher={UMA Project}, 
  author={Lambur, Hart}, 
  year={2024}, 
  month={Jan}
}

@misc{Benligiray_2023, 
  title={Oracle Extractable Value (OEV)}, 
  url={https://medium.com/api3/oracle-extractable-value-oev-13c1b6d53c5b}, 
  journal={Medium}, 
  publisher={API3}, 
  author={Benligiray, Burak}, 
  year={2023}, 
  month={May}
}

@inproceedings{liu2021first,
  title={A first look into defi oracles},
  author={Liu, Bowen and Szalachowski, Pawel and Zhou, Jianying},
  booktitle={2021 IEEE International Conference on Decentralized Applications and Infrastructures (DAPPS)},
  pages={39--48},
  year={2021},
  organization={IEEE}
}

@article{carroll2015robustness,
  title={Robustness and linear contracts},
  author={Carroll, Gabriel},
  journal={American Economic Review},
  volume={105},
  number={2},
  pages={536--563},
  year={2015},
  publisher={American Economic Association 2014 Broadway, Suite 305, Nashville TN 37203}
}

@article{pai2023structural,
  title={Structural Advantages for Integrated Builders in MEV-Boost},
  author={Pai, Mallesh and Resnick, Max},
  journal={arXiv preprint arXiv:2311.09083},
  year={2023}
}

@article{gupta2023centralizing,
  title={The centralizing effects of private order flow on proposer-builder separation},
  author={Gupta, Tivas and Pai, Mallesh M and Resnick, Max},
  journal={arXiv preprint arXiv:2305.19150},
  year={2023}
}

@misc{Charbonneau_2023, 
    title={Suave, anoma, shared sequencers, \& Super Builders}, 
    url={https://dba.mirror.xyz/NTg5FSq1o_YiL_KJrKBOsOkyeiNUPobvZUrLBGceagg}, 
    journal={SUAVE, Anoma, Shared Sequencers, \& Super Builders}, 
    publisher={Mirror}, 
    author={Charbonneau, Jon}, 
    year={2023}, 
    month={4}
}

@incollection{kalagnanam2004auctions,
  title={Auctions, bidding and exchange design},
  author={Kalagnanam, Jayant and Parkes, David C},
  booktitle={Handbook of quantitative supply chain analysis: modeling in the e-business era},
  pages={143--212},
  year={2004},
  publisher={Springer}
}

@misc{the-block-dex,
    title={DeFi Exchange Data and Charts for DEXs, AMMs, and Swaps},
    author={The Block},
    journal={The Block},
    year={2023},
    month={12},
    publisher={The Block},
    url={https://www.theblock.co/data/decentralized-finance/dex-non-custodial}
}

@misc{0xrfq,
	author = {Morales, Fulvia},
	title = {0x API swap now supports 0x Protocol v4},
	howpublished = {\url{https://medium.com/@fulviamorales/0x-api-swap-now-supports-0x-protocol-v4-4c53c732d04f}},
	year = {2021},
        month = {3},
}

@article{o2021electronic,
  title={The electronic evolution of corporate bond dealers},
  author={O`Hara, Maureen and Zhou, Xing Alex},
  journal={Journal of Financial Economics},
  volume={140},
  number={2},
  pages={368--390},
  year={2021},
  publisher={Elsevier}
}

@article{o2010quote,
  title={What is a quote?},
  author={O`Hara, Maureen},
  journal={The Journal of Trading (Retired)},
  volume={5},
  number={2},
  pages={10--16},
  year={2010},
  publisher={Institutional Investor Journals Umbrella}
}

@misc{defillama-dex,
    title={DEXs --- DeFi Llama},
    author={DeFi Llama},
    journal={DeFi Llama},
    year={2023},
    month={12},
    publisher={DeFi Llama},
    url={\url{https://defillama.com/dexs}},
    howpublished={\url{https://defillama.com/dexs}},   
}

@misc{cow-batch-auction, 
  title={The batch auction optimization problem}, 
  howpublished={\url{https://docs.cow.fi/solvers/in-depth-solver-specification/the-batch-auction-optimization-problem}}, 
  journal={The Batch Auction Optimization Problem - CoW Protocol}, 
  author={Leupold, Felix}, 
  year={2021}, 
  month={6}
}

@inproceedings{ramseyer2023speedex,
  title={SPEEDEX: A Scalable, Parallelizable, and Economically Efficient Decentralized $\{$EXchange$\}$},
  author={Ramseyer, Geoffrey and Goel, Ashish and Mazi{\`e}res, David},
  booktitle={20th USENIX Symposium on Networked Systems Design and Implementation (NSDI 23)},
  pages={849--875},
  year={2023}
}

@article{bahrani2023transaction,
  title={Transaction fee mechanism design with active block producers},
  author={Bahrani, Maryam and Garimidi, Pranav and Roughgarden, Tim},
  journal={arXiv preprint arXiv:2307.01686},
  year={2023}
}

@book{krishna2009auction,
  title={Auction theory},
  author={Krishna, Vijay},
  year={2009},
  publisher={Academic press}
}

@book{boyd2004convex,
  title={Convex optimization},
  author={Boyd, Stephen P and Vandenberghe, Lieven},
  year={2004},
  publisher={Cambridge university press}
}

@techreport{sec-prop-605,
    author ={{Securities and Exchange Committee}},
    title = {Proposed Rule: Order Competiton Rule},
    institution = {Securities and Exchange Committee},
    year = {2022},
    month = {Dec},
    version = {Release No. 34-96495},
    howpublished = {\url{https://www.sec.gov/files/rules/proposed/2022/34-96495.pdf}}
}

@article{angel2021gamestonk,
  title={Gamestonk: What happened and what to do about it},
  author={Angel, James},
  journal={Georgetown McDonough School of Business Research Paper},
  number={3782195},
  year={2021}
}

@article{bryzgalova2023retail,
  title={Retail trading in options and the rise of the big three wholesalers},
  author={Bryzgalova, Svetlana and Pavlova, Anna and Sikorskaya, Taisiya},
  journal={The Journal of Finance},
  volume={78},
  number={6},
  pages={3465--3514},
  year={2023},
  publisher={Wiley Online Library}
}

@inproceedings{kulkarni2023routing,
  title={Routing MEV in Constant Function Market Makers},
  author={Kulkarni, Kshitij and Diamandis, Theo and Chitra, Tarun},
  booktitle={International Conference on Web and Internet Economics},
  pages={456--473},
  year={2023},
  organization={Springer}
}

@article{welch2022wisdom,
  title={The wisdom of the Robinhood crowd},
  author={Welch, Ivo},
  journal={The Journal of Finance},
  volume={77},
  number={3},
  pages={1489--1527},
  year={2022},
  publisher={Wiley Online Library}
}

\newpage
\appendix

\section{Exponential value distributions}\label{app:exponential-distribution}

We now derive the left hand side of equation \eqref{eq:threshold} in the case where $F$ is the exponential distribution with rate $\lambda$ (that is, $F(p) = 1-\exp(-\lambda p)$ for $p \geq 0$). We have: 

\begin{align*}
    &\sum_{k=0}^{n} \left({n \choose k} F_C^{k}(\bar{c}) (1- F_C(\bar{c}))^{n-k}\right) S(k)\\
    =& \frac{1}{\lambda} \sum_{k=0}^{n} \left({n \choose k} F_C^{k}(\bar{c}) (1- F_C(\bar{c}))^{n-k}\right) \frac{1}{k+1}\\
    =& \frac{1}{\lambda} \sum_{k=0}^{n} \left( \frac{n!}{k! (n-k)!} F_C^{k}(\bar{c}) (1- F_C(\bar{c}))^{n-k}\right) \frac{1}{k+1}\\
    =& \frac{1}{\lambda} \sum_{k=0}^{n} \left( \frac{n!}{(k+1)! (n-k)!} F_C^{k}(\bar{c}) (1- F_C(\bar{c}))^{n-k}\right)\\
    =&\frac{1}{(n+1)\lambda F_C(\bar{c})} \sum_{k=0}^{n} \left( \frac{(n+1)!}{(k+1)! (n-k)!} F_C^{k+1}(\bar{c}) (1- F_C(\bar{c}))^{n-k}\right)\\
    =&\frac{1}{(n+1)\lambda F_C(\bar{c})} \sum_{k=1}^{n+1} \left( \frac{(n+1)!}{(k)! (n+1-k)!} F_C^{k}(\bar{c}) (1- F_C(\bar{c}))^{n+1 -k}\right)\\
    =& \frac{1}{(n+1)\lambda F_C(\bar{c})} (1- (1-F_C(\bar{c}))^{n+1})
\end{align*}

\section{Uniform value distributions}\label{app:uniform-distribution}

We now derive the left hand side of equation \eqref{eq:threshold} in the case where $F$ is the uniform distribution on $[0,1]$ (that is, $F(p) = p$ for $p \in [0,1]$). We have:
\begin{align*}
        &\sum_{k=0}^{n} \left({n \choose k} F_C^{k}(\bar{c}) (1- F_C(\bar{c}))^{n-k}\right) S(k)\\
        =& \sum_{k=0}^{n} \left({n \choose k} F_C^{k}(\bar{c}) (1- F_C(\bar{c}))^{n-k}\right) \frac{1}{(k+1)(k+2)}\\
        =&  \sum_{k=0}^{n} \left( \frac{n!}{k! (n-k)!} F_C^{k}(\bar{c}) (1- F_C(\bar{c}))^{n-k}\right) \frac{1}{(k+1)(k+2)}\\
    =& \sum_{k=0}^{n} \left( \frac{n!}{(k+2)! (n-k)!} F_C^{k}(\bar{c}) (1- F_C(\bar{c}))^{n-k}\right)\\
    =&\frac{1}{(n+1)(n+2) F_C^2(\bar{c})} \sum_{k=0}^{n} \left( \frac{(n+2)!}{(k+2)! (n-k)!} F_C^{k+2}(\bar{c}) (1- F_C(\bar{c}))^{n-k}\right)\\
    =&\frac{1}{(n+1)(n+2) F_C^2(\bar{c})} \sum_{k=2}^{n+2} \left( \frac{(n+2)!}{(k)! (n-k)!} F_C^{k}(\bar{c}) (1- F_C(\bar{c}))^{n-k}\right)\\
    =& \frac{1}{(n+1)(n+2) F_C^2(\bar{c})} (1- (1-F_C(\bar{c}))^{n+2} - (n+2) F_c(\bar{c}) (1-F_C(\bar{c}))^{n+1})
\end{align*}

\section{Extreme spacing}\label{app:extreme-spacing}
In this section, we derive the equality $(k+1) S(k) = ES(k+1)$ for any CDF $F$ and PDF $f$, where $p^* = 0$. To see this, observe the following chain of equalities:  

    \begin{align*}
        ES(k+1) &= \Expect_{p_i \sim f}[p_{k+1:k+1} - p_{k:k+1}] \\
        &= \int_0^\infty \left((k+1) F(p)^k f(p) - (k+1) k F(p)^{k-1} (1-F(p)) f(p) \right) p dp\\
        &= (k+1) \int_0^\infty F(p)^{k-1} f(p)\left(F(p) -  k (1-F(p))  \right) p dp\\
        &= (k+1) \int_0^\infty F(p)^{k-1} f(p)\left((k+1)F(p) -  k  \right) p dp\\
        &= (k+1) \int_0^\infty f(p) \left((k+1)F(p)^k -  k F(p)^{k-1}  \right) p dp\\
        &= (k+1) \int_0^\infty \frac{d\left(F(p)^{k+1} -   F(p)^{k}  \right)}{dp} pdp\\
        &= (k+1) \int_0^\infty (F(p)^k - F(p)^{k+1}) dp \\
        & = (k+1) \int_{0}^{\infty} F(p)^k (1- F(p)) dp,
    \end{align*}
where the second to last equality follows from integration by parts.

\end{document}